\documentclass[12pt]{article}
\usepackage{graphicx,amssymb,amsmath,empheq}
\usepackage{authblk}
\usepackage{amsthm,bm}
\usepackage{empheq}
\usepackage{subfig} 
\usepackage{soul,physics}
\usepackage{hyperref}
\hypersetup{colorlinks = true, linkcolor=black, citecolor=black, urlcolor=blue}
\usepackage{url}

\theoremstyle{plain}

\usepackage{tikz}
\usepackage{tikz-cd}
\usetikzlibrary{arrows}
\usetikzlibrary{intersections}
\usetikzlibrary{shapes.geometric}
\usetikzlibrary{decorations.pathmorphing, patterns,shapes}
\usetikzlibrary{decorations.markings}

\tikzset{
  mid arrow/.style={postaction={decorate,decoration={
        markings,
        mark=at position .575 with {\arrow{stealth}}
  }}},
  near arrow/.style={postaction={decorate,decoration={
        markings,
        mark=at position .275 with {\arrow{stealth}}
  }}},
  far arrow/.style={postaction={decorate,decoration={
        markings,
        mark=at position .800 with {\arrow{stealth}}
  }}},
  snake arrow/.style={fixed point arithmetic, decorate, decoration={snake,amplitude=2pt, segment length=11pt},postaction={decoration={markings,mark=at position 0.625 with {\arrow{stealth}}},decorate}},
}
\tikzset{
  baseline = -0.5ex,
  wavy/.style = {
    thick,
    decorate,
    decoration={snake,amplitude=2pt,segment length=5pt}},
  swavy/.style = {
    thick,
    decorate,
    decoration={snake,amplitude=1.2pt,segment length=3pt}},
  sdot/.style = {
    circle,
    draw=none,
    fill=black,
    minimum size=2.5pt,
    inner sep=0pt},
  bdot/.style = {
    circle,
    draw=none,
    fill=black,
    minimum size=4pt,
    inner sep=0pt},
  svertex/.style = {
    circle,
    draw=black,
    thick,
    fill=lightgray,
    minimum size=8pt,
    inner sep=1pt},
  bvertex/.style = {
    circle,
    draw=black,
    thick,
    fill=lightgray,
    minimum size=24pt},
  bvertexsmall/.style = {
    circle,
    draw=black,
    thick,
    fill=lightgray,
    minimum size=7pt},
  bvertexnormal/.style = {
    circle,
    draw=black,
    thick,
    fill=lightgray,
    minimum size=16pt},
  mmvertex/.style = {
    circle,
    draw=black,
    thick,
    fill=lightgray,
    minimum size=10pt,
    inner sep=1pt},
  mvertex/.style = {
    circle,
    draw=black,
    thick,
    fill=lightgray,
    minimum size=12pt,
    inner sep=1pt},
  dvertex/.style = {
    circle,
    draw=black,
    thick,
    fill=gray,
    minimum size=25pt}}

\usepackage[nosort]{cite}

\newcommand{\iu}{{i\mkern1mu}}
\newcommand*\diff{\mathop{}\!\mathrm{d}}

\title{
Universal Aspects of High-Temperature Relaxation Dynamics in Random Spin Models
}
\author[1]{Tian-Gang Zhou}
\author[2,3,4]{Wei Zheng}
\author[5,6]{Pengfei Zhang\thanks{pengfeizhang.physics@gmail.com}}
\affil[1]{\normalsize \it Institute for Advanced Study, Tsinghua University, Beijing, 100084, China}
\affil[2]{\normalsize \it Hefei National Research Center for Physical Sciences at the Microscale and School of Physical Sciences, 
University of Science and Technology of China, Hefei 230026, China}
\affil[3]{\normalsize \it CAS Center for Excellence in Quantum Information and Quantum Physics, 
University of Science and Technology of China, Hefei 230026, China}
\affil[4]{\normalsize \it Hefei National Laboratory, 
University of Science and Technology of China, Hefei 230088, China}

\affil[5]{\normalsize \it Department of Physics, Fudan University, Shanghai, 200438, China}
\affil[6]{\normalsize \it Shanghai Qi Zhi Institute, AI Tower, Xuhui District, Shanghai 200232, China}

\date{\today}

\begin{document}
  \maketitle
  
  \begin{abstract}
  Universality is a crucial concept in modern physics, allowing us to capture the essential features of a system's behavior using a small set of parameters. In this work, we unveil universal spin relaxation dynamics in anisotropic random Heisenberg models with infinite-range interactions at high temperatures. Starting from a polarized state, the total magnetization can relax monotonically or decay with long-lived oscillations, determined by the sign of a universal single function $A=-\xi_1^2+\xi_2^2-4\xi_2\xi_3+\xi_3^2$. Here $(\xi_1,\xi_3,\xi_3)$ characterizes the anisotropy of the Heisenberg interaction. Furthermore, the oscillation shows up only for $A>0$, with frequency $\Omega \propto \sqrt{A}$. To validate our theory, we compare it to numerical simulations by solving the Kadanoff-Baym (KB) equation with a melon diagram approximation and the exact diagonalization (ED). The results show our theoretical prediction works in both cases, regardless of a small system size $N=8$ in ED simulations. Our study sheds light on the universal aspect of quantum many-body dynamics beyond low energy limit.
  \end{abstract}
  \tableofcontents

\section{Introduction}

A complete description of realistic many-body systems always contains a large number of parameters. For example, typical solid-state material contains complicated interactions between electrons, phonons, nuclei, and impurities. However, properties that are of physical interest can usually be captured by simple toy models with few parameters. This is a remarkable consequence of universality. The universality states that microscopically different systems can share the same physics at large scales. It usually emerges in long wave length or low-energy limit. For example, phase transitions of many-body systems can be classified into universality classes determined only by the symmetry and dimension of systems~\cite{HalperinTheoryDynamicCritical1977,OdorUniversalityClassesNonequilibrium2004}. Low-energy scattering between atoms can be well described by a single parameter, the scattering length $a_s$, despite details of underlying microscopic interaction potentials \cite{zhai_2021}. Aiming at deepening our understanding of realistic systems, discovering new universalities becomes an important subject in modern  many-body physics. 

\begin{figure}[tb]
    \centering
    \includegraphics[width=0.6\linewidth]{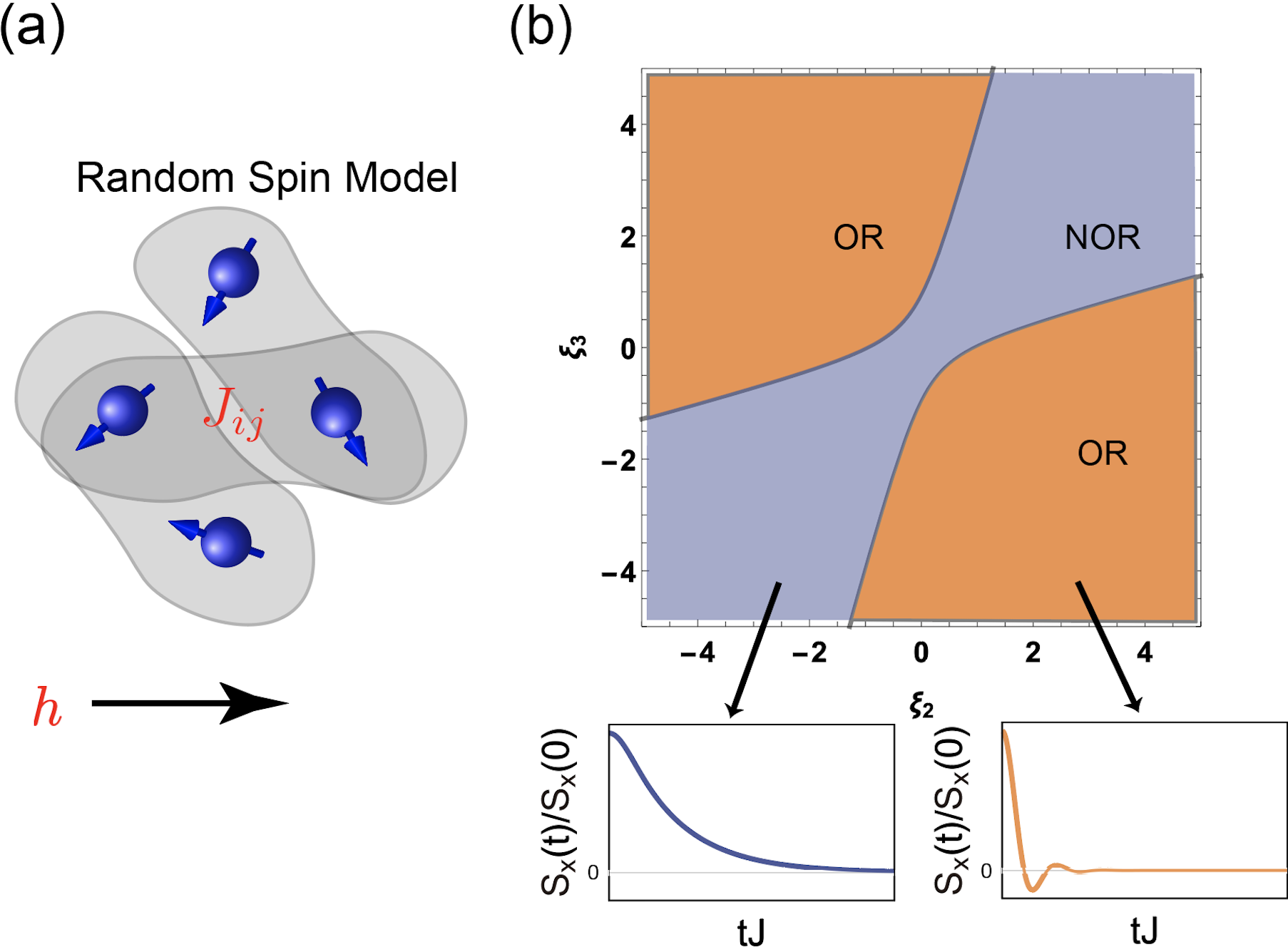}
    \caption{(a). Schematics of the random spin model with random (anisotropic) Heisenberg interactions $J_{ij}$ in the magnetic field $h$. (b). Different dynamical behaviors of the system for different anisotropy parameters $(1, \xi_2,\xi_3)$.The boundary line is determined by $A=-\xi_1^2+\xi_2^2-4\xi_2\xi_3+\xi_3^2=0$, which is symmetric under the reflection along $\xi_2=\pm\xi_3$. OR and NOR denote the oscillating regime and non-oscillating relaxation regime respectively, distinguished by features of the magnetization relaxation process.}
    \label{fig:schemticas}
  \end{figure}

  Recent years have witnessed a great breakthrough in understanding real time dynamics or relaxation in quantum many-body systems both theoretically \cite{palManybodyLocalizationPhase2010,nandkishoreManyBodyLocalizationThermalization2015,abaninColloquiumManybodyLocalization2019,turnerWeakErgodicityBreaking2018,linExactQuantumManyBody2019,iadecolaQuantumManybodyScars2019,linQuantumManybodyScar2020,choiEmergentSUDynamics2019b,serbynQuantumManybodyScars2021a,hoPeriodicOrbitsEntanglement2019,turnerQuantumScarredEigenstates2018,okaPhotovoltaicHallEffect2009,kitagawaTopologicalCharacterizationPeriodically2010,rudnerAnomalousEdgeStates2013a,nathanTopologicalSingularitiesGeneral2015,yaoTopologicalInvariantsFloquet2017,xuTopologicalMicromotionFloquet2022,xuTopologicallyProtectedBoundary2022,liQuantumZenoEffect2018a,liMeasurementdrivenEntanglementTransition2019,skinnerMeasurementInducedPhaseTransitions2019,gullansDynamicalPurificationPhase2020,chanUnitaryprojectiveEntanglementDynamics2019a,jianMeasurementInducedPhaseTransition2021a,jianMeasurementinducedCriticalityRandom2020a,zhangUniversalEntanglementTransitions2022,zhouGeneralizedLindbladMaster2022} and experimentally \cite{Gross:MBL2016,smithManybodyLocalizationQuantum2016,bernienProbingManybodyDynamics2017,LaiManybodyHilbertSpace2023,lindnerFloquetTopologicalInsulator2011,rechtsmanPhotonicFloquetTopological2013,RoushanQuantumInformationPhases2023}. In the previous studies on relaxation, most universal dynamical behaviors emerge in the low temperature or long time scale. That reflects the microscopic details of models are smoothed out in the low energy scale. However at high temperature and short time, it is common believed that most microscopic details are involved in the evolution. Such that the evolution is highly model dependent and hard to observe a universal dynamics. In this work, we unveil that a universal aspect of relaxation dynamics which shows up in an anisotropic Heisenberg model with all-to-all interactions even at high temperatures and short time. The Hamiltonian reads: 
  \begin{equation}
  \hat{H}=\sum\limits_{1\leq i<j\leq N}J_{ij}(\xi_1 \hat{S}^x_i\hat{S}^x_j+ \xi_2 \hat{S}^y_i\hat{S}^y_j+\xi_3\hat{S}^z_i\hat{S}^z_j)-h(t) \sum\limits_{1\leq i \leq N} \hat{S}^x_i. \label{eq:random_spin}
  \end{equation}
  This model with different anisotropy parameters $(\xi_1,\xi_2,\xi_3)$ has been realized in cold molecules \cite{Hazzard:2014bx,Yan:2013fn}, NV centers \cite{Lukin17NV,Lukin18NV}, trapped fermions \cite{smaleObservationTransitionDynamical2019}, Rydberg atoms \cite{Bloch:2012ee,Signoles:2019us}, high spin atoms \cite{gabardosRelaxationCollectiveMagnetization2020}, and solid-state NMR systems \cite{Suter06,Suter07}. A schematic figure is presented in Fig.~\ref{fig:schemticas} (a). Because of random locations or complicated spatial wavefunctions of spin carriers, $J_{ij}$ is usually modeled as independent random Gaussian variables with expectation $\overline{J_{ij}}=\bar{J}/N$ and variance $\overline{\delta J_{ij}^2}=4J^2/N$.
   
We focus on the following protocol: The system is prepared at high temperatures with a polarization field $h(t<0)=h$, which induces a magnetization in the $x$ direction. We then monitor the relaxation of the total magnetization after turning off $h$ suddenly at $t=0$. We find the total magnetization decays either monotonically or with long-lived oscillations, depending on $A=-\xi_1^2+\xi_2^2-4\xi_2\xi_3+\xi_3^2$. The oscillation only appears for $A>0$, in which case the frequency satisfies $\Omega\propto J\sqrt{A}$. Importantly, this phenomenon should be understood as a universal property of the relaxation dynamics since the criterion only contains a specific combination of anisotropic parameters, instead of full details of the microscopic model \eqref{eq:random_spin}. To validate our theoretical prediction, we further perform numerical simulations based on the Kadanoff-Baym (KB) equation with melon-diagram approximations, and the exact diagonalization (ED). Numerical results show the theoretical prediction works in both cases, although we are limited to a small system size $N=8$ in the ED. Our work also provides a novel theoretical framework to analyze the dynamics of randomly interacting quantum spin models.

\section{Kadanoff-Baym equation}
We are interested in the relaxation dynamics of total magnetization. Our theoretical analysis is based on the path-integral approach on the Keldysh contour, as elaborated in \cite{kamenev2023field,stefanucci2013nonequilibrium}. To begin with, we observe that the random spin model can be written in terms of Abrikosov fermion operators $\hat{c}_{i,s}$ with spins $s=\uparrow,\downarrow$ in the single occupation subspace. Explicitly, we have $\hat{S}^\alpha_i=\sum_{ss'}\frac{1}{2}\hat{c}^\dagger_{i,s}(\sigma^\alpha)_{s s^\prime} \hat{c}_{i,s^\prime}$, where $\alpha=x,y,z$ and $\sigma^\alpha$ denote the corresponding Pauli matrices. Since the Hamiltonian \eqref{eq:random_spin} exhibits $\pi$ rotation symmetries along the $x$ axis, the total magnetization can only be along the $x$ axis. We thus introduce $m(t)\equiv\langle\hat S^x(t)\rangle$. Since the total magnetization is always along the $x$ direction, the magnetization can be computed by real-time Green's functions of fermion operators:
\begin{equation}\label{eq:magnetization}
m(t)=-i G^{>}_{\uparrow\downarrow}(t,t)=-i G^{<}_{\uparrow\downarrow}(t,t),
\end{equation} 
where we have defined 
\begin{equation}
\begin{aligned}\label{eq:Gless_great}
G^>_{ss'}(t_1,t_2)&\equiv-i  \sum_l\left<c_{l,s}(t_1)c_{l,s'}^\dagger(t_2)\right>/N,\\
G^<_{ss'}(t_1,t_2)&\equiv i  \sum_l\left<c_{l,s'}^\dagger(t_2)c_{l,s}(t_1)\right>/N.
\end{aligned}
\end{equation}

The relaxation dynamics of $m(t)$ can then be computed once we obtain the Green's functions $G^\gtrless(t_1,t_2)$. It is known that the evolution of $G^\gtrless(t_1,t_2)$ is governed by the Kadanoff-Baym equation, which can be derived by the Schwinger-Dyson equation on the Schwinger-Keldysh contour.
\begin{equation}
  \begin{aligned}\label{eq:KBeq}
    i\partial_{t_1}&G^\gtrless+ \frac{1}{2}h_{\text{eff}}(t_1)\sigma^xG^\gtrless=\Sigma^R\circ G^\gtrless+\Sigma^\gtrless\circ G^A, \\
    -i\partial_{t_2}&G^\gtrless+\frac{1}{2}h_{\text{eff}}(t_2)G^\gtrless\sigma^x=G^R\circ\Sigma^\gtrless+G^\gtrless\circ \Sigma^A.
  \end{aligned}
\end{equation}
Here we have introduced self-energies $\Sigma^{\gtrless}$ and $\Sigma^{R/A}$. We define the operation $\circ$ for functions with two time variables as $f\circ g\equiv\int dt_3~f(t_1,t_3)g(t_3,t_2)$. The retarded and advanced Green's functions $G^{R/A}$ are related to $G^{\gtrless}$ by $G^{R/A}=\pm \Theta\left(\pm t_{12}\right)\left(G^>-G^<\right)$, where $\Theta(t)$ is the Heaviside step function. Similar relations work for self-energies $\Sigma^{R/A}$. $h_{\text{eff}}(t)=h(t)+\bar{J}m(t)$ is the effective magnetic field, which includes the mean-field contribution from $\bar{J}$. For $t<0$, the system is prepared in thermal equilibrium. Consequently, we have $G^{\gtrless}(t_1,t_2)=G^{\gtrless}_\beta(t_{12})$ for $t_1,t_2<0$. For either $t_1>0$ or $t_2>0$, the Green's functions evolve due to the quantum quench and should be obtained by solving Eq. \eqref{eq:KBeq} after the self-energy is specified. 

The approximation comes in when we try to relate the self-energies to Green's functions. After transforming into the Abrikosov fermion representation, the random Heisenberg interaction takes the form of random fermion scatterings. Interestingly, such random interaction terms is a close analog of the celebrated complex Sachdev-Ye-Kitaev (SYK) model \cite{kitaevalexei2015,maldacenaRemarksSachdevYeKitaevModel2016a,songStronglyCorrelatedMetal2017a,davisonThermoelectricTransportDisordered2017,guoTransportChaosLattice2019b,Gu:2020it,zhouDisconnectingTraversableWormhole2021}. Motivated by this observation, here we make the melon diagram approximation for the fermion self-energy. A formal argument to control errors is to generalize the Hamiltonian \eqref{eq:random_spin} into large-$M$ spins, as in the seminal work by Sachdev and Ye \cite{sachdevGaplessSpinfluidGround1993}. We promote the original model by adding an additional $M$ indices as
\begin{equation}\label{eqn:hhh}
  \hat{H}= \frac{1}{\sqrt{M}} \sum_{i<j,\alpha \gamma} J_{ij}\xi^\alpha\hat{T}^{\alpha,\gamma}_i\hat{T}^{\alpha,\gamma}_j-h\sum_i\hat{S}^x_i.
  \end{equation}
  where we have introduced $\hat{T}^{\alpha,\gamma}_i=\frac{1}{2}\sum_{s_i,m_i}\hat{c}^{\dagger}_{i,s_1,m_1}(\sigma^\alpha)_{s_1s_2}(T^\gamma)_{m_1m_2}\hat{c}_{i,s_2,m_2}$ with $\gamma\in\{1,2,...,M^2-1\}$ labeling the generators of the $SU(M)$ group. It is known that they satisfy the completeness relation $\sum_\gamma T^\gamma_{m_1m_2}T^\gamma_{m_3m_4}=\delta_{m_1m_4}\delta_{m_2m_3}-\frac{1}{M}\delta_{m_1m_2}\delta_{m_3m_4}$. The external field $h$ only couples to the $SU(2)$ part. The constrain is also promoted as $\sum_{s,m}\hat{c}^{\dagger}_{i,s,m}\hat{c}_{i,s,m}=M$. Firstly, we take the imaginary time approach in the large-$N$ and large-$M$ limits. The constrain is satisfied automatically due to the particle-hole symmetry, and the self-energy can be obtained by the melon diagrams as in \cite{sachdevGaplessSpinfluidGround1993}. Finally, this leads to
\begin{equation}\label{eq:selfreal}
  \begin{aligned}
    \Sigma^{\gtrless}&(t_1,t_2)=\frac{J^2}{4}\sum_{\alpha,\alpha^\prime}\xi_\alpha \xi_{\alpha^\prime} \sigma^{\alpha^\prime} G^{\gtrless}(t_1,t_2)\sigma^\alpha\text{Tr}\left[\sigma^{\alpha^\prime} G^{\gtrless}(t_1,t_2)\sigma^\alpha G^{\lessgtr}(t_2,t_1)\right],
  \end{aligned}
\end{equation}
Here, we omit spin indices for conciseness. We have introduced the anisotropy vector $\bm{\xi}=(\xi_1,\xi_2,\xi_3)$. The melon diagram approximation may fail in the low-temperature limit if the system exhibits spin glass orders \cite{Baldwin:2019dki}. In this work, we avoid this problem by focusing on the high-temperature regime with $\beta J\ll 1$. Combining Eq. \eqref{eq:magnetization}, \eqref{eq:KBeq}, and \eqref{eq:selfreal} leads to a set of closed equations which determines the relaxation of the magnetization. For later convenience, we also provide matrix elements of Eq. \eqref{eq:selfreal} explicitly after using the symmetry of Green's function in Appendix~\ref{suppsec:sym}. Leaving the details in the Appendix.~\ref{suppsec:selfenergy}, we have
\begin{equation}\label{eqS:full_self_energy}
  \Sigma^{\gtrless}
  = \frac{J^2}{2}  \begin{pmatrix}
    -\bm{\xi}^2 G^{\gtrless}_{\uparrow\uparrow}(t_1,t_2)^3 + A G^{\gtrless}_{\uparrow\uparrow}(t_1,t_2) G^{\gtrless}_{\uparrow\downarrow}(t_1,t_2)^2& 
    A G^{\gtrless}_{\uparrow\uparrow}(t_1,t_2)^2 G^{\gtrless}_{\uparrow\downarrow}(t_1,t_2) + \bm{\xi}^2 G^{\gtrless}_{\uparrow\downarrow}(t_1,t_2)^3 \\
    A G^{\gtrless}_{\uparrow\uparrow}(t_1,t_2)^2 G^{\gtrless}_{\uparrow\downarrow}(t_1,t_2) + \bm{\xi}^2 G^{\gtrless}_{\uparrow\downarrow}(t_1,t_2)^3& 
    -\bm{\xi}^2 G^{\gtrless}_{\uparrow\uparrow}(t_1,t_2)^3 + A G^{\gtrless}_{\uparrow\uparrow}(t_1,t_2) G^{\gtrless}_{\uparrow\downarrow}(t_1,t_2)^2 \\
  \end{pmatrix},
\end{equation} 
with $A=-\xi_1^2+\xi_2^2-4\xi_2\xi_3+\xi_3^2$ and $\bm{\xi}^2 \equiv \xi_1^2+\xi_2^2+\xi_3^2$.
\begin{figure}[tb]
    \centering
    \includegraphics[width=0.45\linewidth]{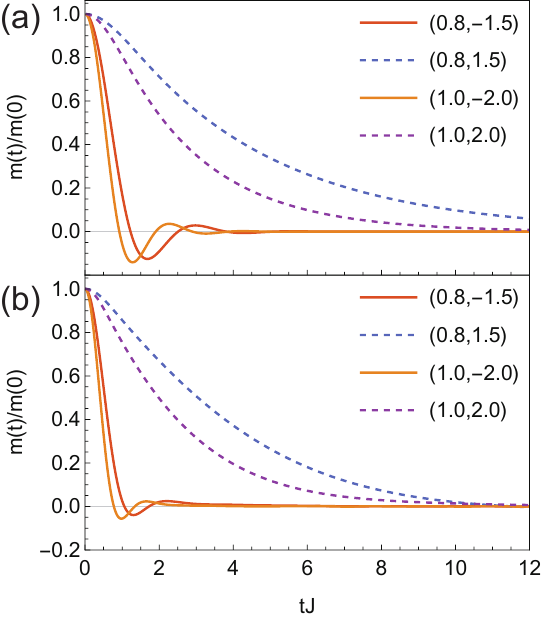}
    \caption{The numerical result for the evolution of the magnetization $m(t)$ by numerically solving: (a) the Kadanoff-Baym equation \eqref{eq:KBeq} and (b) exact diagonalization with system size $N=8$. Initially, the system is in thermal equilibrium with $\beta J=0.04$, $\bar{J}=0$ and $h/J=10$. We take $\xi_1=1$ and consider four different anisotropy parameters $(\xi_2,\xi_3)=(0.8,-1.5)$, $(0.8,1.5)$, $(1,-2)$, and $(1,2)$, which corresponds to $A=6.69$, $-2.91$, $12$, and $-4$. The results show that the relaxation of $m(t)$ is monotonic/oscillating if $A<0$/$A>0$. These two numerical results match each other to good precision despite a small $N$.}
    \label{fig:numerics_magnetization}
\end{figure}

Typical numerical results for $m(t)$ obtained by two methods are shown in Fig.~\ref{fig:numerics_magnetization}. Here we consider examples with $\xi_1=1$ and $(\xi_2,\xi_3)=(0.8,-1.5)$, $(0.8,1.5)$, $(1,-2)$, and $(1,2)$. We set the initial temperature $\beta J=0.04$, the polarization field $h/J=10$ and $\bar{J}=0$. In the long-time limit, the system exhibits the quantum thermalization to the thermal ensemble with $h=0$. In this case, $\pi$ rotations along $y$ or $z$ also become the symmetry of the Hamiltonian, which makes $m(\infty)=0$. According to the relaxation process, different anisotropy parameters can be divided into two groups, under which $m(t)$ relaxes monotonically (for $(\xi_2,\xi_3)=(0.8,1.5)$ and $(1,2)$) or with long-lived oscillations (for $(\xi_2,\xi_3)=(0.8,-1.5)$ and $(1,-2)$). Furthermore, we numerically checked that the presence of the oscillation is stable against deformations of parameters. As a result, we propose the Hamiltonian \eqref{eq:random_spin} with different $(\xi_1,\xi_2,\xi_3)$ can be separated in parameter regimes with oscillating relaxation (OR) versus non-oscillating relaxation (NOR), as shown in Fig.~\ref{fig:schemticas}. In Appendix \ref{A}, we verify that the difference in the dynamical behavior can not be detected in equilibrium via spin susceptibility.

\section{Oscillation versus monotonic decay}
After evolving for a long time, the total magnetization, as well as off-diagonal components of Green's functions, becomes very small. Consequently, we can perform a linearized analysis of the KB equation to reveal the mechanism for the oscillation and determine the criterion for different dynamical behaviors. A differential equation that governs the long-time evolution of the magnetization can be derived following a few steps:
\vspace{10pt}

\textbf{Step 1.--} The linearized analysis can be largely simplified after the Keldysh rotation. We introduce the standard Keldysh Green's function of fermions as $G^K=G^>+G^<$. The total magnetization can be expressed as its off-diagonal component:
\begin{equation}\label{eq:mtGK}
m(t)=-iG^K_{\uparrow\downarrow}(t,t)/2.
\end{equation}
We can further combine equations in \eqref{eq:KBeq} to derive the equation for $G^K$.
On the Keldysh contour, the Schwinger-Dyson equation reads 
\begin{equation}\label{eq:Keldysh_consist}
  \begin{pmatrix}
      G^R & G^K \\
        0   & G^A
  \end{pmatrix}^{-1}= \begin{pmatrix}
      G_0^R & 0 \\
        0   & G_0^A
  \end{pmatrix} - \begin{pmatrix}
      \Sigma^R & \Sigma^K \\
        0   & \Sigma^A
  \end{pmatrix}.
\end{equation}
Here, we have $\Sigma^K=\Sigma^>+\Sigma^<$. Taking the Keldysh component of Eq. \ref{eq:Keldysh_consist}, we find
\begin{equation}\label{eq:SD_GK_Keldysh}
  G^K=G^R \circ \Sigma^K \circ G^A.
\end{equation} 

\textbf{Step 2.--} We linearize Eq.~\eqref{eq:SD_GK_Keldysh} around the equilibrium solution in the long-time limit after the quantum thermalization.  We expand $G^{a}(t_1,t_2)=G^{a,\beta_f}(t_{12})+\delta G^{a}(t_1,t_2)$, where $G^{a,\beta_f}(t)$ is the equilibrium Green's function on the final state. Leaving the details in the appendix.~\ref{suppsec:selfenergy}, the off-diagonal element of \eqref{eq:SD_GK_Keldysh} reads 
\begin{equation}\label{eq:SD_GK_Keldyshlinear}
\begin{aligned}
  \delta G^K_{\uparrow \downarrow}&=G^{R,\beta_f}_{\uparrow \uparrow} \circ \delta \Sigma^K_{\uparrow \downarrow}\circ G^{A,\beta_f}_{\uparrow \uparrow},\\ \delta \Sigma^K_{\uparrow \downarrow}&= \frac{1}{4} J^2 A \left((G^{>,\beta_f}_{\uparrow\uparrow})^2  + (G^{<,\beta_f}_{\uparrow\uparrow})^2 \right) \delta G^{K}_{\uparrow\downarrow}.
  \end{aligned}
\end{equation} 
where we have used the fact that the equilibrium state contains no magnetization, and is approximately at infinite temperature. The second fact leads to $\Sigma^{K,\beta_f} \approx 0$. Since $G^{K,\beta_f}_{\uparrow \downarrow}=0$, Eq. \eqref{eq:mtGK} is equivalent to $m(t)=-i\delta G^K_{\uparrow \downarrow}(t,t)/2$.

\textbf{Step 3.--} To proceed, we need to obtain approximations for $G^{a,\beta_f}_{\uparrow\uparrow}$. In thermal equilibrium with $h=0$, the self-energies \eqref{eq:selfreal} can be simplified as
\begin{equation}\label{eq:latetime}
\Sigma^{\gtrless,\beta_f}_{ss'}(t)=-\frac{J^2\bm{\xi}^2}{2}G^{\gtrless,\beta_f}_{ss}(t)^3\delta_{ss'},
\end{equation}
where we have used $G^{>,\beta_f}_{ss}(t)=-G^{<,\beta_f}_{ss}(-t)$ due to the particle-hole symmetry. Eq.~\eqref{eq:latetime} then matches the self-energy of the Majorana SYK$_4$ model with effective coupling constant $J|\bm{\xi}|/\sqrt{2}$. It is known that at high temperatures $\beta J\ll1$, the SYK model can be described by weakly interacting quasi-particles \cite{zhangObstacleSubAdSHolography2021}. Taking the Lorentzian approximation, we have
\begin{equation}\label{eq:highT_GF}
\begin{aligned}
  G_{\uparrow\uparrow}^{R/A,\beta_f}(t)\approx\mp i\Theta(\pm t)e^{-\Gamma |t|/2},\ \ \ \ \ \ 
  G_{\uparrow\uparrow}^{\gtrless,\beta_f}(t)\approx\mp ie^{-\Gamma |t|/2}/2.
  \end{aligned}
\end{equation}
with quasi-particle decay rate $\Gamma \propto J$. 

\textbf{Step 4.--} Using the high-temperature solution, we get
\begin{equation}\label{eq:conv}
  \begin{split}
    \delta G^K_{\uparrow \downarrow}(t_1, t_4) &= \int \diff t_2 \diff t_3 \  G^R_{\uparrow \uparrow}(t_1, t_2)  \left( -\frac{1}{8}J^2 A e^{-|t_2-t_3|\Gamma } \delta G^K_{\uparrow \downarrow}(t_2, t_3) \right)  G^A_{\downarrow \downarrow}(t_3, t_4)  \\
    &= -\frac{1}{8}J^2 A \int \diff t_2 \diff t_3 
    e^{-\frac{\Gamma}{2}(t_1-t_2)} \Theta(t_1-t_2)
    e^{-|t_2-t_3|\Gamma } \delta G^K_{\uparrow \downarrow}(t_2, t_3)
    e^{\frac{\Gamma}{2}(t_3-t_4)} \Theta(-t_3+t_4).
    \\
  \end{split}
\end{equation}
Multiply $e^{\frac{\Gamma}{2} t_1}$ and take $\partial_{t_1}$ on the Eq.~\eqref{eq:conv}, which gives
\begin{equation}\label{eq:conv_partial_t1}
  \begin{split}
    \partial_{t_1} \left[e^{\frac{\Gamma}{2} t_1} \delta G^K_{\uparrow \downarrow}(t_1, t_4) \right]
    &= -\frac{1}{8}J^2 A \int \diff t_2 \diff t_3 
    e^{\frac{\Gamma}{2}t_2} \delta(t_1-t_2)
    e^{-|t_2-t_3|\Gamma } \delta G^K_{\uparrow \downarrow}(t_2, t_3)
    e^{\frac{\Gamma}{2}(t_3-t_4)} \Theta(-t_3+t_4)
    \\
  &= -\frac{1}{8}J^2 A \int  \diff t_3 
    e^{\frac{\Gamma}{2}t_1} 
    e^{-|t_1-t_3|\Gamma } \delta G^K_{\uparrow \downarrow}(t_1, t_3)
    e^{\frac{\Gamma}{2}(t_3-t_4)} \Theta(-t_3+t_4).
    \\
  \end{split}
\end{equation}
Then multiply $e^{\frac{\Gamma}{2} t_4}$ and again take $\partial_{t_4}$ on both sides of Eq.~\eqref{eq:conv_partial_t1}, which leads to a differential equation
\begin{equation}\label{eq:SD_GK_diff}
\left(\partial_{t_1}+\frac{\Gamma}{2}\right)\left(\partial_{t_2}+\frac{\Gamma}{2}\right)\delta G^K_{\uparrow \downarrow}=-\frac{A}{8}J^2
   e^{-\Gamma|t_{12}|} \delta G^K_{\uparrow \downarrow}.
\end{equation}
\vspace{10pt}

Eq. \eqref{eq:SD_GK_diff} is the starting point for analyzing the relaxation dynamics. Since it is invariant under time translations, we separate out the center-of-mass time dependence by introducing $\delta G^K_{\uparrow \downarrow}(t_1,t_2)=\text{Re}~e^{-\lambda\frac{t_1+t_2}{2}}\varphi(t_{12})$. The relaxation is oscillatory only if $\lambda$ is complex. Interestingly, $\varphi(t_{12})$ then satisfies the 1D Schr$\ddot{\text{o}}$dinger equation
\begin{equation}\label{eq:schoinger}
-\frac{(\Gamma-\lambda)^2}{4}\varphi(t_{12}) =-\partial_{t_{12}}^2 \varphi(t) + \frac{A}{8} J^2e^{-\Gamma |t_{12}|}\varphi(t_{12}),
\end{equation} 
where $-\frac{(\Gamma-\lambda)^2}{4}$ plays the role of the energy $E$ and $\frac{A}{8} J^2e^{-\Gamma |t_{12}|}$ plays the role of potential $V$. Eq.~\eqref{eq:schoinger} suggests the boundary line between the oscillating regime and the non-oscillating regime is at $A=0$: For $A<0$, the potential energy is negative. It is known that in 1D any attractive potential exhibits at least one bound state. Denoting the energy of the ground state as $-|E_0|$, we can solve $\lambda=\Gamma-2\sqrt{|E_0|}$, which is real. Consequently, we expect the magnetization relaxes monotonically. For $A>0$, the potential is repulsive. The eigenstates of the \eqref{eq:schoinger} are scattering modes with continuous positive energy $E$. We find $\lambda=\Gamma\pm2i\sqrt{E}$, which is complex. This leads to oscillations in the relaxation process. 

\begin{figure}[tb]
    \centering
    \includegraphics[width=0.6\linewidth]{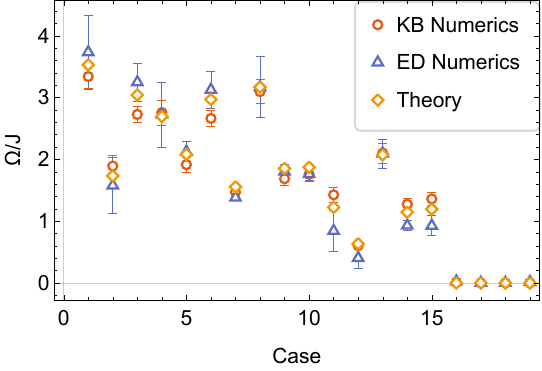}
    \caption{A comparison between the theoretical prediction $\Omega_{Th}/J=c_0 \sqrt{A}$ and numerical simulations. Here we choose $c_0=1$. In each case, we randomly choose the anisotropic parameter $(\xi_1,\xi_2,\xi_3)$. Initially, the system is also in thermal equilibrium with $\beta J=0.04$, $\bar{J}=0$, and $h/J=10$. The numerical data are obtained by fitting numerics based on the KB equation and ED method. The error bars correspond to standard deviations when concerning the different fitting regions.}
    \label{fig:FreqNumerics}
  \end{figure}

To further determine the typical oscillation frequency $\Omega$, we need to determine the typical energy $E$ that contributes to the quench dynamics. According to Eq. \eqref{eq:mtGK}, the magnetization probes the decay of the wave function at $t_{12}=0$, where the potential energy is $\sim AJ^2$. For $E\ll AJ^2$, the eigenstate has exponentially small weight near $t_{12}=0$. As a result, the corresponding contribution to $m(t)$ can be neglected. We can approximate
\begin{equation}
m(t)\sim \int_{AJ^2} dE~c(E)e^{-\Gamma t-2i\sqrt{E}t}.
\end{equation}
Here $c(E)$ is some smooth function determined by the initial condition. We then expect $\Omega \approx c_0 \sqrt{A}J$, with some $O(1)$ constant $c_0$ which does not depend on parameters in the Hamiltonian \eqref{eq:random_spin} and should be extracted using numerics. Interestingly, the result predicts the oscillation period $T=2\pi/\Omega$ diverges as we approach $A=0$, which can be viewed as an analog of the divergence of the correlation length in traditional phase transition described by order parameters.

We comment that our results unveil the universality of relaxation dynamics in random spin models. Although the microscopic model in Eq.~\eqref{eq:random_spin} contains several parameters, the criterion for the different relaxation behaviors, as well as the oscillation frequency, only depends on a specific combination $A$. This is a direct analog of universality in the scattering theory, where for a complicated potential, the low-energy scattering problem can only depend on a specific combination of microscopic parameters, which is the scattering length.

We further compare our prediction of the oscillation frequency $\Omega\approx c_0 \sqrt{A} J$ to numerical results. We obtain $\Omega$ in numerics by fitting $m(t)=m_0 \cos(\Omega t+\theta)e^{-\Gamma t} + m_{\text{offset}}$. Here $m_0$ is the amplitude, $\theta$ is the phase, $\Gamma$ is the quasi-particle decay rate, and $m_{\text{offset}}$ is the offset which is significant in the finite $N$ ED numerics. The fitting particularly focuses on the matching in the small $m(t)$ region. Hence the detailed fitting region and the error bars caused by such ambiguity are left to the Appendix.~\ref{App:fig3}. The results are shown in Fig.~\ref{fig:FreqNumerics}. We randomly choose the anisotropic parameters, and the first 15 cases correspond to $A>0$, and the last 4 cases to $A<0$ (see Appendix.~\ref{App:fig3}). Among the $A>0$ cases, the mean ratio between the numerical data and the polynomial $A$ reads $\overline{\Omega_{\text{KB}}/ (J\sqrt{A})} = 0.995 \pm 0.018$ and $\overline{\Omega_{\text{ED}}/ (J\sqrt{A})} = 0.94 \pm 0.04$. Therefore, we set $c_0=1$ for theoretical predictions in Fig.~\ref{fig:FreqNumerics}. 
 Although the error bars for ED numerics are significantly larger than KB numerics since the calculation is based on the finite $N=8$ system, we find the theoretical prediction of the oscillation frequency almost matches the KB results and the ED results, up to the error bars. From Fig.~\ref{fig:FreqNumerics}, most notably, the OR and NOR relaxation are sharply distinguished by the $A>0$ or $A<0$ criterion, which is perfectly aligned with our theoretical analysis.


\section{Discussions} 
In this work, we show that the random Heisenberg model with all-to-all interactions exhibits universal relaxation dynamics governed by a single parameter $A=-\xi_1^2+\xi_2^2-4\xi_2\xi_3+\xi_3^2$. Unlike traditional examples where the universality emerges in the low-energy limit, here the universal physics appears at high temperatures. For $A<0$, the magnetization decays monotonically after we turn off the polarization field. For $A>0$, long-lived oscillation appears during the relaxation process, with a frequency $\Omega \propto J\sqrt{A}$. Our theoretical analysis is based on the path-integral approach on the Keldysh contour, which is verified by comparing our theory to numerical simulations by solving the KB equation or the ED.  

We remark that quantum coherence is essential for the existence of the oscillating relaxation regime. As an example, if we spoil the coherence by considering time-dependent random interactions instead of static interactions, the magnetization is expected to decay monotonically: After replacing $J_{ij}$ with Brownian variables $J_{ij}(t)$, Eq.~\eqref{eq:SD_GK_diff} is replaced by
\begin{equation}\label{eq:SD_GK_diffBrownian}
\left(\partial_{t}+{\Gamma}\right)\delta G^K_{\uparrow \downarrow}(t,t)=-\frac{AJ}{8} \delta G^K_{\uparrow \downarrow}(t,t),
\end{equation}
as derived in the Appendix \ref{B}. This results in $m(t)\sim e^{-(\Gamma +AJ/8)t}$ with a simple exponential decay, on contrary to the existence of different dynamical behaviors in the static case.

We also point out that amazingly our criteria $A>0$ for the oscillation regime matches the criteria proposed in \cite{zhouDisconnectingTraversableWormhole2021} for the presence of the instability towards the formation of wormholes with $\xi_1=\xi_2=1$. However, the analysis in \cite{zhouDisconnectingTraversableWormhole2021} focuses on the low-temperature regime, while in this work we focus on high temperatures. This makes it difficult to establish a direct relationship between the two theoretical analyses. It would be interesting to explore whether there is some form of duality between the high-temperature and low-temperature limits. Given that the wormhole phase is non-chaotic, it would also be intriguing to study the out-of-time-order correlator or the operator size distribution in regimes with different dynamical behaviors. Experimentally, our results can be readily verified through quantum quench experiments in NMR systems.

\vspace{5pt}
\textit{Note Added.} Universal behaviors of auto-correlation function related to the quench dynamics discussed here, including oscillatory versus non-oscillatory behavior, have been related to Lanczos coefficients computed for determining the Krylov complexity in Ref.~\cite{Zhang:2023wtr}.

\section*{Acknowledgment}
We are especially grateful to the invaluable discussions with Hui Zhai, whose advice is indispensable for the whole work. We thank Riqiang Fu, Yuchen Li, Xinhua Peng, Xiao-Liang Qi and Ren Zhang for their helpful discussions.

\appendix 

\section{Susceptibility in thermal equilibrium}\label{A}

\begin{figure}
  \centering
  \includegraphics[width=0.8\linewidth]{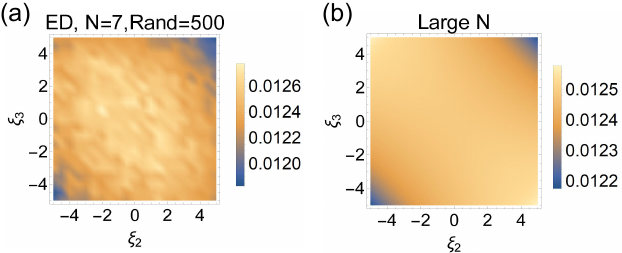}
  \caption{(a) The susceptibility of the exact diagonalization numerics. We choose site number $N=7$, and random realizations to be $500$. (b) The susceptibility of the large-$N$ numerics. In both cases, to obtain the susceptibility, we choose the equilibrium thermal state with temperature $T/J=10.0$, and external magnetic field with $h/J=0.05$ in $x$ direction. We take the region of anisotropic parameter $\xi_2, \xi_3$ to be $\left[-5, 5\right]$, and the discretization step is $\Delta \xi = 0.5$.}
  \label{fig:susc}
\end{figure}

As we discussed in the main text, we calculate the equilibrium susceptibility. We take a small external magnetic field in the $x$ direction in  the equilibrium thermal state. The finite difference susceptibility is defined as $\chi = \langle \hat{S}_x \rangle_h /h$, where $\langle \hat{S}_x \rangle_h$ means the thermal average in the external magnetic field $h$. First, we find the exact diagonalization and large-$N$ Kadanoff-Baym results in Fig.~\ref{fig:susc} agree well with each other. Second, equilibrium susceptibility in Fig.~\ref{fig:susc} is highly in contrast with the criterion for the relaxation dynamics $A$ in Fig.~\ref{fig:schemticas} (b). There is no appreciable distinction between $A>0$ and $A<0$ region correspondingly in the plot of the equilibrium. Hence, it reveals the significance of our dynamical framework.

\section{Analysis for Brownian interactions}\label{B}

To check the effect of coherence, we place $J_{ij}$ in the Hamiltonian Eq.~\eqref{eqn:hhh} with Brownian variables $J_{ij}(t)$, where $\overline{J_{ij}(t_1) J_{ij}(t_2)} = 4J^2/N \delta(t_1 -t_2)$.The self energy is replaced by
\begin{equation}\label{eq:selfE_brownian}
\begin{aligned}
  \Sigma^{\gtrless}&(t_1,t_2)
  = \frac{1}{2} J^2 \delta(t_1-t_2) \\
  &\begin{pmatrix}
    -\bm{\xi}^2 G^{\gtrless}_{\uparrow\uparrow}(t_1,t_2)^3 + A G^{\gtrless}_{\uparrow\uparrow}(t_1,t_2) G^{\gtrless}_{\uparrow\downarrow}(t_1,t_2)^2& 
    A G^{\gtrless}_{\uparrow\uparrow}(t_1,t_2)^2 G^{\gtrless}_{\uparrow\downarrow}(t_1,t_2) + \bm{\xi}^2 G^{\gtrless}_{\uparrow\downarrow}(t_1,t_2)^3 \\
    A G^{\gtrless}_{\uparrow\uparrow}(t_1,t_2)^2 G^{\gtrless}_{\uparrow\downarrow}(t_1,t_2) + \bm{\xi}^2 G^{\gtrless}_{\uparrow\downarrow}(t_1,t_2)^3& 
    -\bm{\xi}^2 G^{\gtrless}_{\uparrow\uparrow}(t_1,t_2)^3 + A G^{\gtrless}_{\uparrow\uparrow}(t_1,t_2) G^{\gtrless}_{\uparrow\downarrow}(t_1,t_2)^2 \\
  \end{pmatrix},
  \end{aligned}
\end{equation}
Therefore, following the same steps, the linearization of the Schwinger-Dyson equation $G^K=G^R \circ \Sigma^K \circ G^A$ leads to
\begin{equation}\label{eq:SD_GK_Keldyshlinear_brownian}
  \begin{aligned}
  \delta G^K_{\uparrow \downarrow}(t_1, t_1)&=\left[G^{R,\beta_f}_{\uparrow \uparrow} \circ \delta \Sigma^K_{\uparrow \downarrow}\circ G^{A,\beta_f}_{\uparrow \uparrow}\right](t_1, t_1),\\ 
  \delta \Sigma^K_{\uparrow \downarrow}(t_1,t_2)&= \frac{1}{4} J^2 A \delta(t_1-t_2) \left(G^{>,\beta_f}_{\uparrow\uparrow}(t_1,t_1)^2  + G^{<,\beta_f}_{\uparrow\uparrow}(t_1,t_1)^2 \right) \delta G^{K}_{\uparrow\downarrow}(t_1,t_1).
  \end{aligned}
\end{equation} 
To form close equation groups, we set the time argument of the first equation to be $(t_1, t_1)$, and then we study the equal time perturbation of the Green's function $\delta G^K_{\uparrow \downarrow}(t, t)$.

The equilibrium Green's function still has the form of Eq.~\eqref{eq:highT_GF} with Lorentz approximation.
Using such a solution, we get
\begin{equation}\label{eq:convBrownian}
  \begin{split}
    \delta G^K_{\uparrow \downarrow}(t_1, t_1) &= \int \diff t_2\  G^R_{\uparrow \uparrow}(t_1, t_2)  \left( -\frac{1}{8}J^2 A e^{-|t_2-t_2|\Gamma } \delta G^K_{\uparrow \downarrow}(t_2, t_2) \right)  G^A_{\downarrow \downarrow}(t_2, t_1)  \\
    &= -\frac{1}{8}J^2 A \int \diff t_2 
    e^{-\frac{\Gamma}{2}(t_1-t_2)} \Theta(t_1-t_2)
     \delta G^K_{\uparrow \downarrow}(t_2, t_2)
    e^{\frac{\Gamma}{2}(t_2-t_1)} \Theta(-t_2+t_1).
    \\
    &= -\frac{1}{8}J^2 A \int \diff t_2 
    e^{-\Gamma(t_1-t_2)} \Theta(t_1-t_2)
    \delta G^K_{\uparrow \downarrow}(t_2, t_2).
    \\
  \end{split}
\end{equation}
Multiply $e^{\Gamma t_1}$ and take $\partial_{t_1}$ on the Eq.~\eqref{eq:convBrownian}, which gives
\begin{equation}\label{eq:convBrownian_partial_t1}
  \begin{split}
    \partial_{t_1} \left[e^{\Gamma t_1} \delta G^K_{\uparrow \downarrow}(t_1, t_1) \right]
    &= -\frac{1}{8}J^2 A \int \diff t_2 \ 
    e^{\Gamma t_2} \delta(t_1-t_2)
    \delta G^K_{\uparrow \downarrow}(t_2, t_2).
    \\
  \end{split}
\end{equation}
Finally leads to
\begin{equation}\label{eq:SD_GK_diffBrownian}
  \left(\partial_{t}+{\Gamma}\right)\delta G^K_{\uparrow \downarrow}(t,t)=-\frac{AJ}{8} \delta G^K_{\uparrow \downarrow}(t,t).
\end{equation}
Solving this differential equation leads to $\delta G^K_{\uparrow\downarrow}(t,t) \sim m(t)\sim e^{-(\Gamma +AJ/8)t}$ with a simple exponential decay without any oscillation for arbitrary anisotropy parameters.

\section{Symmetry on Green's function}\label{suppsec:sym}
The greater Green's function in Eq.~\eqref{eq:Gless_great} can be explicitly expressed in $2\times 2$ matrix form.
\begin{equation}\label{eq:GF_matrix}
	\begin{split}
		G^>_{s_1 s_2}(t_1,t_2) = -i \langle \hat{c}_{s_1}(t_1) \hat{c}^{\dagger}_{s_2}(t_2) \rangle  &= \begin{pmatrix}
			G^>_{\uparrow \uparrow}(t_1,t_2) & G^>_{\uparrow \downarrow}(t_1,t_2) \\
			G^>_{\downarrow \uparrow}(t_1,t_2) & G^>_{\downarrow \downarrow}(t_1,t_2) \\
		\end{pmatrix}_{s_1 s_2} \\
    \end{split}
\end{equation}

There are two symmetries crucial for the later simplification. The first can be regarded as $\pi$ rotation of axis $x$.
\begin{equation}\label{eq:Sym1}
	\begin{split}
		\hat{c}_{s_1} &\to \sum_{s'} \left(\iu \sigma^x \right)_{s_1 s'}\hat{c}_{s'} \\
		\hat{c}_{s_1}^{\dagger} &\to \sum_{s'} \hat{c}_{s'}^{\dagger} \left(-\iu \sigma^x \right)_{s' s_1}. \\
	\end{split}
\end{equation}
With symmetry in Eq.~\eqref{eq:Sym1}, $\left\{\hat{S}^{x},\hat{S}^{y},\hat{S}^{z}\right\}$ is mapped to $\left\{\hat{S}^{x},-\hat{S}^{y},-\hat{S}^{z}\right\}$, and therefore keeps Eq.~\eqref{eq:random_spin} invariant. As a consequence, the symmetry of Green's function leads to
\begin{equation}\label{eq:Sym1_GF}
	\begin{split}
		G^>_{s_1 s_2}(t_1,t_2) 
		&\rightarrow \sum_{s' s''} \sigma^x_{s_1 s'} G^>_{s' s''}(t_1,t_2) \sigma^x_{s'' s_2} \\
		&= \begin{pmatrix}
			G^>_{\downarrow \downarrow}(t_1,t_2) & G^>_{\downarrow \uparrow}(t_1,t_2) \\
			G^>_{\uparrow \downarrow}(t_1,t_2) & G^>_{\uparrow \uparrow}(t_1,t_2) \\
		\end{pmatrix}_{s_1 s_2} \\
	\end{split}
\end{equation}
The second symmetry is combined with particle-hole symmetry and rotation, which reads as
\begin{equation}\label{eq:Sym2}
	\begin{split}
		\hat{c}_{s_1} &\to \sum_{s'} \left(\iu \sigma^y \right)_{s_1 s'}\hat{c}_{s'}^{\dagger} \\
		\hat{c}_{s_1}^{\dagger} &\to \sum_{s'} \hat{c}_{s'} \left(-\iu \sigma^y \right)_{s' s_1}. \\
	\end{split}
\end{equation}
Similarly, with symmetry in Eq.~\eqref{eq:Sym2}, $\left\{\hat{S}^{x},\hat{S}^{y},\hat{S}^{z}\right\}$ is mapped to $\left\{\hat{S}^{x},-\hat{S}^{y},\hat{S}^{z}\right\}$, and therefore keeps Eq.~\eqref{eq:random_spin} invariant. Also, the symmetry of Green's function leads to
\begin{equation}\label{eq:Sym2_GF}
	\begin{split}
		G^>_{s_1 s_2}(t_1,t_2)
		&\rightarrow -\sum_{s' s''} \sigma^y_{s_1 s'} G^<_{s'' s'}(t_2,t_1) \sigma^y_{s'' s_2} \\
		&= \begin{pmatrix}
			-G^<_{\uparrow \uparrow}(t_2,t_1) & G^<_{\uparrow \downarrow}(t_2,t_1) \\
			G^<_{\uparrow \downarrow}(t_2,t_1) & -G^<_{\uparrow \uparrow}(t_2,t_1) \\
		\end{pmatrix}_{s_1 s_2}, \\
	\end{split}
\end{equation}
where the second line to the third line uses the symmetry obtained in Eq.~\eqref{eq:Sym1_GF}. Finally, we can exchange $>$ and $<$ symbols in Eq.~\eqref{eq:Sym1_GF} and Eq.~\eqref{eq:Sym1_GF} to obtain another two symmetry in terms of Green's function.

\section{Derivation of the self-energy}\label{suppsec:selfenergy}
\subsection{Simplification of Eq.~\eqref{eq:selfreal}}
Starting from Eq.~\eqref{eq:selfreal}, we derive the corresponding analytical formula. We first consider the trace part $\text{Tr}\left[\sigma^{\alpha^\prime} G^{\gtrless}(t_1,t_2)\sigma^\alpha G^{\lessgtr}(t_2,t_1)\right]$. In the basis of $\alpha',\alpha=\left\{x,y,z\right\}$, direct calculation of the trace part reads
\begin{equation}\label{eq:tracepart}
	\begin{split}
		&\text{Tr}\left[\sigma^{\alpha^\prime} G^{\gtrless}(t_1,t_2)\sigma^\alpha G^{\lessgtr}(t_2,t_1)\right] \\
		=&-\text{Tr}\left[\sigma^y \sigma^{\alpha^\prime} G^{\gtrless}(t_1,t_2)\sigma^\alpha \sigma^y  G^{\gtrless}(t_1,t_2)\right] \\
		=&\begin{pmatrix}
			-2 G^{\gtrless}_{\uparrow\uparrow}(t_1,t_2)^2 + 2 G^{\gtrless}_{\uparrow\downarrow}(t_1,t_2)^2 & 0 & 0 &\\
			0 & -2 G^{\gtrless}_{\uparrow\uparrow}(t_1,t_2)^2 - 2 G^{\gtrless}_{\uparrow\downarrow}(t_1,t_2)^2 & 4\iu G^{\gtrless}_{\uparrow\uparrow}(t_1,t_2) G^{\gtrless}_{\uparrow\downarrow}(t_1,t_2) &\\
			0 & -4\iu G^{\gtrless}_{\uparrow\uparrow}(t_1,t_2) G^{\gtrless}_{\uparrow\downarrow}(t_1,t_2) & -2 G^{\gtrless}_{\uparrow\uparrow}(t_1,t_2)^2 - 2 G^{\gtrless}_{\uparrow\downarrow}(t_1,t_2)^2 &\\
		\end{pmatrix}_{\alpha' \alpha}. \\
	\end{split}
\end{equation}
The second line applies the symmetry Eq.~\eqref{eq:Sym2_GF} and implicitly use Eq.~\eqref{eq:Sym1_GF} to ensure $\left(G^{\gtrless}\right)^T=G^{\gtrless}$. The off-diagonal matrix elements in the first row and column correspond to the $x$ direction, which is nontrivially disappeared. Since the initial external field is in the $x$ direction, such anisotropic is exactly reflected in the components of self-energy by applying the symmetry constraints.

The self-energy composites different components $\alpha, \alpha'$ which represents the internal spin interaction. Therefore we show each non-zero contribution in Eq.~\eqref{eq:selfreal}.

\paragraph{$\alpha'=x,\alpha=x$}
\begin{equation}\label{eqS:xx}
	\frac{1}{2} J^2 \xi_1^2 
		\begin{pmatrix}
			G^{\gtrless}_{\uparrow\uparrow}(t_1,t_2) & G^{\gtrless}_{\uparrow\downarrow}(t_1,t_2) \\
			G^{\gtrless}_{\uparrow\downarrow}(t_1,t_2) & G^{\gtrless}_{\uparrow\uparrow}(t_1,t_2) \\
		\end{pmatrix}
	\left(- G^{\gtrless}_{\uparrow\uparrow}(t_1,t_2)^2 +  G^{\gtrless}_{\uparrow\downarrow}(t_1,t_2)^2\right)
\end{equation}

\paragraph{$\alpha'=y,\alpha=y$}
\begin{equation}\label{eqS:yy}
	\frac{1}{2} J^2 \xi_2^2 
	\begin{pmatrix}
		-G^{\gtrless}_{\uparrow\uparrow}(t_1,t_2) & G^{\gtrless}_{\uparrow\downarrow}(t_1,t_2) \\
		G^{\gtrless}_{\uparrow\downarrow}(t_1,t_2) & -G^{\gtrless}_{\uparrow\uparrow}(t_1,t_2) \\
	\end{pmatrix}
	\left( G^{\gtrless}_{\uparrow\uparrow}(t_1,t_2)^2 +  G^{\gtrless}_{\uparrow\downarrow}(t_1,t_2)^2\right)
\end{equation}

\paragraph{$\alpha'=z,\alpha=z$}
\begin{equation}\label{eqS:zz}
	\frac{1}{2} J^2 \xi_3^2 
	\begin{pmatrix}
		-G^{\gtrless}_{\uparrow\uparrow}(t_1,t_2) & G^{\gtrless}_{\uparrow\downarrow}(t_1,t_2) \\
		G^{\gtrless}_{\uparrow\downarrow}(t_1,t_2) & -G^{\gtrless}_{\uparrow\uparrow}(t_1,t_2) \\
	\end{pmatrix}
	\left( G^{\gtrless}_{\uparrow\uparrow}(t_1,t_2)^2 +  G^{\gtrless}_{\uparrow\downarrow}(t_1,t_2)^2\right)
\end{equation}

\paragraph{$\alpha'=y,\alpha=z$}
\begin{equation}\label{eqS:yz}
	 J^2 \xi_2 \xi_3 
	\begin{pmatrix}
		G^{\gtrless}_{\uparrow\downarrow}(t_1,t_2) & -G^{\gtrless}_{\uparrow\uparrow}(t_1,t_2) \\
		-G^{\gtrless}_{\uparrow\uparrow}(t_1,t_2) & G^{\gtrless}_{\uparrow\downarrow}(t_1,t_2) \\
	\end{pmatrix}
	\left( G^{\gtrless}_{\uparrow\uparrow}(t_1,t_2)G^{\gtrless}_{\uparrow\downarrow}(t_1,t_2)\right)
\end{equation}

\paragraph{$\alpha'=z,\alpha=y$}
\begin{equation}\label{eqS:zy}
	J^2 \xi_2 \xi_3 
	\begin{pmatrix}
		G^{\gtrless}_{\uparrow\downarrow}(t_1,t_2) & -G^{\gtrless}_{\uparrow\uparrow}(t_1,t_2) \\
		-G^{\gtrless}_{\uparrow\uparrow}(t_1,t_2) & G^{\gtrless}_{\uparrow\downarrow}(t_1,t_2) \\
	\end{pmatrix}
	\left( G^{\gtrless}_{\uparrow\uparrow}(t_1,t_2)G^{\gtrless}_{\uparrow\downarrow}(t_1,t_2)\right)
\end{equation}

We finally arrive at the full self-energy by collecting all these terms  
\begin{equation}\label{eqS:full_self_energy}
	\Sigma^{\gtrless}(t_1,t_2)
	= \frac{J^2 }{2} \begin{pmatrix}
		-\bm{\xi}^2 G^{\gtrless}_{\uparrow\uparrow}(t_1,t_2)^3 + A G^{\gtrless}_{\uparrow\uparrow}(t_1,t_2) G^{\gtrless}_{\uparrow\downarrow}(t_1,t_2)^2& 
		A G^{\gtrless}_{\uparrow\uparrow}(t_1,t_2)^2 G^{\gtrless}_{\uparrow\downarrow}(t_1,t_2) + \bm{\xi}^2 G^{\gtrless}_{\uparrow\downarrow}(t_1,t_2)^3 \\
		A G^{\gtrless}_{\uparrow\uparrow}(t_1,t_2)^2 G^{\gtrless}_{\uparrow\downarrow}(t_1,t_2) + \bm{\xi}^2 G^{\gtrless}_{\uparrow\downarrow}(t_1,t_2)^3& 
		-\bm{\xi}^2 G^{\gtrless}_{\uparrow\uparrow}(t_1,t_2)^3 + A G^{\gtrless}_{\uparrow\uparrow}(t_1,t_2) G^{\gtrless}_{\uparrow\downarrow}(t_1,t_2)^2 \\
	\end{pmatrix},
\end{equation}
where the polynomials $\bm{\xi}^2=\xi_1^2+\xi_2^2+\xi_3^2$ and $A=-\xi_1^2+\xi_2^2-4\xi_2\xi_3+\xi_3^2$.

\subsection{Perturbation}
With perturbation on the off-diagonal term on the Eq.~\eqref{eq:SD_GK_Keldysh}, in principle we have
\begin{equation}\label{eq:linearization}
     \delta G^K_{\uparrow \downarrow}=\sum_{s_1, s_2} \underbrace{ G^{R,\beta_f}_{\uparrow s_1} \circ \delta \Sigma^K_{s_1 s_2}\circ G^{A,\beta_f}_{s_2 \downarrow} }_{\text{Eq.~\eqref{eq:linearization}(a)}} 
  + \underbrace{\delta G^{R}_{\uparrow s_1} \circ  \Sigma^{K,\beta_f}_{s_1 s_2}\circ G^{A,\beta_f}_{s_2 \downarrow} }_{\text{Eq.~\eqref{eq:linearization}(b)}}
  + \underbrace{G^{R,\beta_f}_{\uparrow s_1} \circ  \Sigma^{K,\beta_f}_{s_1 s_2}\circ \delta G^{A}_{s_2 \downarrow} }_{\text{Eq.~\eqref{eq:linearization}(c)}}.
\end{equation}
For Eq.~\eqref{eq:linearization}(a), all off-diagonal components of $G^{R/A,\beta_f}$ are zero, and consequently we only take $s_1 = \uparrow$ and $s_2=\downarrow$. For Eq.~\eqref{eq:linearization}(b), the vanishing off-diagonal component of $G^{A,\beta_f}, \Sigma^{K,\beta_f}$ requires $s_2=\downarrow$ and then $s_1 = \downarrow$. However, in the infinite high-temperature region we have $\Sigma^{K,\beta_f}_{\downarrow \downarrow}=0$ as a result of the fluctuation-dissipation theorem\cite{kamenev2023field,stefanucci2013nonequilibrium}. In the following content, we can also verify this with the specific ansatz for equilibrium Green's function in Eq.~\eqref{eq:latetime},~\eqref{eq:highT_GF}. Therefore Eq.~\eqref{eq:linearization}(b) vanishes, and we can show Eq.~\eqref{eq:linearization}(c) vanishes with similar arguments. Combining all the argument, only the off-diagonal term in Eq.~\eqref{eq:linearization}(a) survives and finally leads to Eq.~\eqref{eq:SD_GK_Keldyshlinear}.

To obtain $\delta \Sigma^K_{s_1 s_2}$ with $s_1 = \uparrow$ and $s_2=\downarrow$, we can use the relation $\Sigma^K=\Sigma^>+\Sigma^<$ and the off-diagonal term in Eq.~\eqref{eqS:full_self_energy}. Near the equilibrium solution, $G^{\gtrless}_{\uparrow\uparrow}(t_1,t_2)^2 G^{\gtrless}_{\uparrow\downarrow}(t_1,t_2) \approx G^{\gtrless,\beta_f}_{\uparrow\uparrow}(t_1,t_2)^2 \delta G^{\gtrless}_{\uparrow\downarrow}(t_1,t_2)$ and $ G^{\gtrless}_{\uparrow\downarrow}(t_1,t_2)^3\approx  \delta G^{\gtrless}_{\uparrow\downarrow}(t_1,t_2)^3$. We will drop the third order contribution in $\delta G_{\uparrow \downarrow}^{\gtrless}$ and only keep the linear order. Finally, it leads to the result
\begin{equation}
    \begin{split}
        \delta \Sigma^K_{\uparrow \downarrow} &= \frac{1}{2}J^2 A G^{>,\beta_f}_{\uparrow\uparrow}(t_1,t_2)^2 \delta G^{>}_{\uparrow\downarrow}(t_1,t_2) + \frac{1}{2}J^2 A G^{<,\beta_f}_{\uparrow\uparrow}(t_1,t_2)^2 \delta G^{<}_{\uparrow\downarrow}(t_1,t_2) \\
    &= \frac{1}{4}J^2 A \left(G^{>,\beta_f}_{\uparrow\uparrow}(t_1,t_2)^2 + G^{<,\beta_f}_{\uparrow\uparrow}(t_1,t_2)^2 \right) \delta G^{K}_{\uparrow\downarrow}(t_1,t_2). \\
    \end{split}
\end{equation}

\section{Estimation of frequency and error bar in Fig.~\ref{fig:FreqNumerics}}\label{App:fig3}

\begin{figure}
	\centering
	\includegraphics[width=1.0\linewidth]{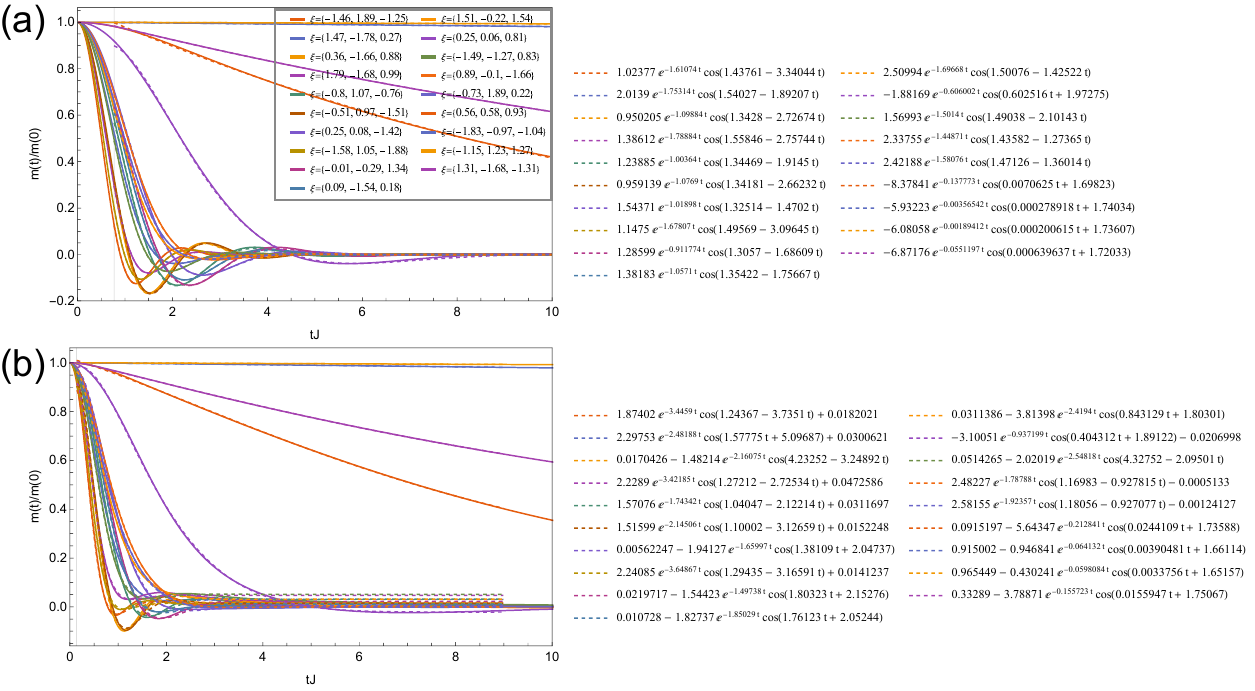}
	\caption{Fitting of (a) the large-$N$ dynamics; (b) the exact diagonalization dynamics corresponding to the main text fig.~\ref{fig:FreqNumerics}. The solid line is the numerical data, and the dashed line is the fitting curve. The inset legend lists each anisotropic parameter corresponding to the 19 cases in the main text. Besides, the right panel shows the fitting functions for each case.}
	\label{fig:figLargeN_ED_supp_fit}
\end{figure}

Fig.~\ref{fig:FreqNumerics} in the main text shows the estimation of frequency and error bars with random anisotropic parameters. Here we show the detailed fitting result in fig.~\ref{fig:figLargeN_ED_supp_fit}, where the initial temperature and external magnetic field is  $\beta J=0.04$, $\bar{J}=0$ and $h/J=10$.

The choice of the fitting region is naturally uncertain. First, the theoretical prediction of the oscillation frequency acquires the assumption of small $m(t)$, but the small $m(t)$ time region is not uniquely defined. Secondly, especially for ED numerics, finite system size leads to untrustable results in the late time limit. Therefore it is reasonable only to consider the early time when fitting the ED numerics, which also introduces uncertainty of fitting region.

The error bars of the fitting frequency arise from such uncertainty. To quantify such error, we separately consider two different numerical approaches.
\begin{itemize}
	\item For the large-$N$ data, we choose the time region begins at different values: $t_{\text{begin}}J=0,0.05,0.1,\cdots,1.2$, and ends at the same point $t_{\text{end}}J=10.0$. We fit these frequencies and take the standard deviation as the error bars.
	\item For the ED data, we choose the time region begins at different values: $t_{\text{begin}}J=0,0.08$,
    $0.16,\cdots,0.4$, and ends at also different points $t_{\text{end}}J=3.0,4.0,\cdots9.0$. We take a combination of these beginning and ending points and take the standard deviation among these frequencies as the error bars.
\end{itemize}
We summarize the detailed fitting region in table.~\ref{tab:parameter_cases}. Notice that for the ED approach, we need a larger ending fitting time for the small frequency cases, but in other cases, we choose a smaller ending fitting time to avoid the finite $N$ effect.

\begin{table}[t]
	\centering
 \resizebox{\textwidth}{!}{%
		\begin{tabular}{|c|c|c|c|c|c|c|c|c|c|c|c|c|c|c|c|c|c|c|c|}
		\hline
		Cases & 1     & 2     & 3     & 4     & 5     & 6     & 7     & 8     & 9     & 10    & 11    & 12    & 13    & 14    & 15    & 16    & 17    & 18    & 19 \\
		\hline
		$\xi_1$ & -1.46 & 1.47  & 0.36  & 1.79  & -0.8  & -0.51 & 0.25  & -1.58 & -0.01 & 0.09  & 1.51  & 0.25  & -1.49 & 0.89  & -0.73 & 0.56  & -1.83 & -1.15 & 1.31 \\
		\hline
		$\xi_2$ & 1.89  & -1.78 & -1.66 & -1.68 & 1.07  & 0.97  & 0.08  & 1.05  & -0.29 & -1.54 & -0.22 & 0.06  & -1.27 & -0.1  & 1.89  & 0.58  & -0.97 & 1.23  & -1.68 \\
		\hline
		$\xi_3$ & -1.25 & 0.27  & 0.88  & 0.99  & -0.76 & -1.51 & -1.42 & -1.88 & 1.34  & 0.18  & 1.54  & 0.81  & 0.83  & -1.66 & 0.22  & 0.93  & -1.04 & 1.27  & -1.31 \\
		\hline
		$\Omega_{\text{KB}}$ & 3.34  & 1.89  & 2.73  & 2.76  & 1.91  & 2.66  & 1.47  & 3.10  & 1.69  & 1.76  & 1.43  & 0.60  & 2.10  & 1.27  & 1.36  & 0.01  & 0.00  & 0.00  & 0.00  \\
		\hline
		$\sigma_{\Omega_{\text{KB}}}$ & 0.19  & 0.17  & 0.13  & 0.20  & 0.12  & 0.13  & 0.10  & 0.20  & 0.10  & 0.12  & 0.12  & 0.03  & 0.16  & 0.10  & 0.11  & 0.01  & 0.00  & 0.00  & 0.01  \\
		\hline
		$\Omega_{\text{ED}}$ & 3.74  & 1.58  & 3.25  & 2.73  & 2.12  & 3.13  & 1.38  & 3.17  & 1.80  & 1.76  & 0.84  & 0.40  & 2.10  & 0.93  & 0.93  & 0.02  & 0.00  & 0.00  & 0.02  \\
		\hline
		$\sigma_{\Omega_{\text{ED}}}$ & 0.59  & 0.46  & 0.31  & 0.52  & 0.17  & 0.30  & 0.05  & 0.49  & 0.11  & 0.10  & 0.33  & 0.16  & 0.23  & 0.08  & 0.16  & 0.05  & 0.04  & 0.04  & 0.04  \\
		\hline
		$\Omega_{\text{Th}}$ & 3.53  & 1.73  & 3.04  & 2.69  & 2.08  & 2.97  & 1.55  & 3.17  & 1.85  & 1.87  & 1.22  & 0.63  & 2.07  & 1.14  & 1.19  & 0.00  & 0.00  & 0.00  & 0.00  \\
		\hline
		$t_{\text{begin,KB}}J$ & 0.75  & 0.75  & 0.75  & 0.75  & 0.75  & 0.75  & 0.75  & 0.75  & 0.75  & 0.75  & 0.75  & 0.75  & 0.75  & 0.75  & 0.75  & 0.75  & 0.75  & 0.75  & 0.75 \\
		\hline
		$t_{\text{end,KB}}J$ & 10    & 10    & 10    & 10    & 10    & 10    & 10    & 10    & 10    & 10    & 10    & 10    & 10    & 10    & 10    & 10    & 10    & 10    & 10 \\
		\hline
		$t_{\text{begin,ED}}J$ & 0.15  & 0.15  & 0.15  & 0.15  & 0.15  & 0.15  & 0.15  & 0.15  & 0.15  & 0.15  & 0.15  & 0.15  & 0.15  & 0.15  & 0.15  & 0.15  & 0.15  & 0.15  & 0.15 \\
		\hline
		$t_{\text{end,ED}}J$ & 3.5   & 3.5   & 3.5   & 3.5   & 3.5   & 3.5   & 3.5   & 3.5   & 3.5   & 3.5   & 3.5   & 9     & 3.5   & 9     & 9     & 9     & 9     & 9     & 9 \\
		\hline
	\end{tabular}%
 }
	\caption{The detailed parameters and result for each case. $(\xi_1, \xi_2, \xi_3)$ means anisotropic parameters. $\Omega_{\text{KB}}, \Omega_{\text{ED}}, \Omega_{\text{Th}}$ are the fitting frequencies from Kadanoff-Baym numerics, ED numerics, and theoretical prediction. $\sigma_{\Omega_{\text{KB}}}, \sigma_{\Omega_{\text{ED}}}$ are the standard deviations for KB and ED numerics. $[t_{\text{begin,KB}}J, t_{\text{end,KB}}J]$ and $[t_{\text{begin,ED}}J, t_{\text{end,ED}}J]$ are the fit region which lead to the $\Omega_{\text{KB}}, \Omega_{\text{ED}}$ numerics data of the main text fig.~\ref{fig:FreqNumerics}. }
	\label{tab:parameter_cases}%
\end{table}%

\bibliography{ref.bib}

\providecommand{\href}[2]{#2}\begingroup\raggedright\begin{thebibliography}{10}

\bibitem{HalperinTheoryDynamicCritical1977}
P.~C. Hohenberg and B.~I. Halperin, \emph{Theory of dynamic critical
  phenomena}, \href{https://doi.org/10.1103/RevModPhys.49.435}{\emph{Rev. Mod.
  Phys.} {\bfseries 49} (July, 1977) 435--479}.

\bibitem{OdorUniversalityClassesNonequilibrium2004}
G.~{\'O}dor, \emph{Universality classes in nonequilibrium lattice systems},
  \href{https://doi.org/10.1103/RevModPhys.76.663}{\emph{Rev. Mod. Phys.}
  {\bfseries 76} (Aug., 2004) 663--724}.

\bibitem{zhai_2021}
H.~Zhai, \emph{Ultracold Atomic Physics}.
\newblock Cambridge University Press, 2021,
  \href{https://doi.org/10.1017/9781108595216}{10.1017/9781108595216}.

\bibitem{palManybodyLocalizationPhase2010}
A.~Pal and D.~A. Huse, \emph{Many-body localization phase transition},
  \href{https://doi.org/10.1103/PhysRevB.82.174411}{\emph{Phys. Rev. B}
  {\bfseries 82} (2010) 174411}.

\bibitem{nandkishoreManyBodyLocalizationThermalization2015}
R.~Nandkishore and D.~A. Huse, \emph{Many-{{Body Localization}} and
  {{Thermalization}} in {{Quantum Statistical Mechanics}}},
  \href{https://doi.org/10.1146/annurev-conmatphys-031214-014726}{\emph{Annu.
  Rev. Condens. Matter Phys.} {\bfseries 6} (2015) 15--38}.

\bibitem{abaninColloquiumManybodyLocalization2019}
D.~A. Abanin, E.~Altman, I.~Bloch and M.~Serbyn, \emph{Colloquium: Many-body
  localization, thermalization, and entanglement},
  \href{https://doi.org/10.1103/RevModPhys.91.021001}{\emph{Rev. Mod. Phys.}
  {\bfseries 91} (May, 2019) 021001}.

\bibitem{turnerWeakErgodicityBreaking2018}
C.~J. Turner, A.~A. Michailidis, D.~A. Abanin, M.~Serbyn and Z.~Papi{\'c},
  \emph{Weak ergodicity breaking from quantum many-body scars},
  \href{https://doi.org/10.1038/s41567-018-0137-5}{\emph{Nat. Phys.} {\bfseries
  14} (2018) 745--749}.

\bibitem{linExactQuantumManyBody2019}
C.-J. Lin and O.~I. Motrunich, \emph{Exact {{Quantum Many-Body Scar States}} in
  the {{Rydberg-Blockaded Atom Chain}}},
  \href{https://doi.org/10.1103/PhysRevLett.122.173401}{\emph{Phys. Rev. Lett.}
  {\bfseries 122} (2019) 173401}.

\bibitem{iadecolaQuantumManybodyScars2019}
T.~Iadecola, M.~Schecter and S.~Xu, \emph{Quantum many-body scars from magnon
  condensation}, \href{https://doi.org/10.1103/PhysRevB.100.184312}{\emph{Phys.
  Rev. B} {\bfseries 100} (2019) 184312}.

\bibitem{linQuantumManybodyScar2020}
C.-J. Lin, V.~Calvera and T.~H. Hsieh, \emph{Quantum many-body scar states in
  two-dimensional {{Rydberg}} atom arrays},
  \href{https://doi.org/10.1103/PhysRevB.101.220304}{\emph{Phys. Rev. B}
  {\bfseries 101} (2020) 220304}.

\bibitem{choiEmergentSUDynamics2019b}
S.~Choi, C.~J. Turner, H.~Pichler, W.~W. Ho, A.~A. Michailidis, Z.~Papi{\'c}
  et~al., \emph{Emergent {{SU}}(2) {{Dynamics}} and {{Perfect Quantum Many-Body
  Scars}}}, \href{https://doi.org/10.1103/PhysRevLett.122.220603}{\emph{Phys.
  Rev. Lett.} {\bfseries 122} (2019) 220603}.

\bibitem{serbynQuantumManybodyScars2021a}
M.~Serbyn, D.~A. Abanin and Z.~Papi{\'c}, \emph{Quantum many-body scars and
  weak breaking of ergodicity},
  \href{https://doi.org/10.1038/s41567-021-01230-2}{\emph{Nat. Phys.}
  {\bfseries 17} (2021) 675--685}.

\bibitem{hoPeriodicOrbitsEntanglement2019}
W.~W. Ho, S.~Choi, H.~Pichler and M.~D. Lukin, \emph{Periodic {{Orbits}},
  {{Entanglement}}, and {{Quantum Many-Body Scars}} in {{Constrained Models}}:
  {{Matrix Product State Approach}}},
  \href{https://doi.org/10.1103/PhysRevLett.122.040603}{\emph{Phys. Rev. Lett.}
  {\bfseries 122} (2019) 040603}.

\bibitem{turnerQuantumScarredEigenstates2018}
C.~J. Turner, A.~A. Michailidis, D.~A. Abanin, M.~Serbyn and Z.~Papi{\'c},
  \emph{Quantum scarred eigenstates in a {{Rydberg}} atom chain:
  {{Entanglement}}, breakdown of thermalization, and stability to
  perturbations}, \href{https://doi.org/10.1103/PhysRevB.98.155134}{\emph{Phys.
  Rev. B} {\bfseries 98} (2018) 155134}.

\bibitem{okaPhotovoltaicHallEffect2009}
T.~Oka and H.~Aoki, \emph{Photovoltaic hall effect in graphene},
  \href{https://doi.org/10.1103/PhysRevB.79.081406}{\emph{Phys. Rev. B}
  {\bfseries 79} (Feb, 2009) 081406}.

\bibitem{kitagawaTopologicalCharacterizationPeriodically2010}
T.~Kitagawa, E.~Berg, M.~Rudner and E.~Demler, \emph{Topological
  characterization of periodically driven quantum systems},
  \href{https://doi.org/10.1103/PhysRevB.82.235114}{\emph{Phys. Rev. B}
  {\bfseries 82} (Dec, 2010) 235114}.

\bibitem{rudnerAnomalousEdgeStates2013a}
M.~S. Rudner, N.~H. Lindner, E.~Berg and M.~Levin, \emph{Anomalous {{Edge
  States}} and the {{Bulk-Edge Correspondence}} for {{Periodically Driven
  Two-Dimensional Systems}}},
  \href{https://doi.org/10.1103/PhysRevX.3.031005}{\emph{Phys. Rev. X}
  {\bfseries 3} (2013) 031005}.

\bibitem{nathanTopologicalSingularitiesGeneral2015}
F.~Nathan and M.~S. Rudner, \emph{Topological singularities and the general
  classification of {{Floquet}}\textendash{{Bloch}} systems},
  \href{https://doi.org/10.1088/1367-2630/17/12/125014}{\emph{New J. Phys.}
  {\bfseries 17} (2015) 125014}.

\bibitem{yaoTopologicalInvariantsFloquet2017}
S.~Yao, Z.~Yan and Z.~Wang, \emph{Topological invariants of floquet systems:
  General formulation, special properties, and floquet topological defects},
  \href{https://doi.org/10.1103/PhysRevB.96.195303}{\emph{Phys. Rev. B}
  {\bfseries 96} (Nov, 2017) 195303}.

\bibitem{xuTopologicalMicromotionFloquet2022}
P.~Xu, W.~Zheng and H.~Zhai, \emph{Topological micromotion of {{Floquet}}
  quantum systems},
  \href{https://doi.org/10.1103/PhysRevB.105.045139}{\emph{Phys. Rev. B}
  {\bfseries 105} (2022) 045139}.

\bibitem{xuTopologicallyProtectedBoundary2022}
P.~Xu and T.-S. Deng, \emph{Topologically protected boundary discrete time
  crystal for a solvable model},
  \href{https://arxiv.org/abs/2210.15222}{{\ttfamily 2210.15222}}.

\bibitem{liQuantumZenoEffect2018a}
Y.~Li, X.~Chen and M.~P.~A. Fisher, \emph{Quantum zeno effect and the many-body
  entanglement transition},
  \href{https://doi.org/10.1103/PhysRevB.98.205136}{\emph{Phys. Rev. B}
  {\bfseries 98} (Nov, 2018) 205136}.

\bibitem{liMeasurementdrivenEntanglementTransition2019}
Y.~Li, X.~Chen and M.~P.~A. Fisher, \emph{Measurement-driven entanglement
  transition in hybrid quantum circuits},
  \href{https://doi.org/10.1103/PhysRevB.100.134306}{\emph{Phys. Rev. B}
  {\bfseries 100} (Oct, 2019) 134306}.

\bibitem{skinnerMeasurementInducedPhaseTransitions2019}
B.~Skinner, J.~Ruhman and A.~Nahum, \emph{Measurement-induced phase transitions
  in the dynamics of entanglement},
  \href{https://doi.org/10.1103/PhysRevX.9.031009}{\emph{Phys. Rev. X}
  {\bfseries 9} (Jul, 2019) 031009}.

\bibitem{gullansDynamicalPurificationPhase2020}
M.~J. Gullans and D.~A. Huse, \emph{Dynamical purification phase transition
  induced by quantum measurements},
  \href{https://doi.org/10.1103/PhysRevX.10.041020}{\emph{Phys. Rev. X}
  {\bfseries 10} (Oct, 2020) 041020}.

\bibitem{chanUnitaryprojectiveEntanglementDynamics2019a}
A.~Chan, R.~M. Nandkishore, M.~Pretko and G.~Smith, \emph{Unitary-projective
  entanglement dynamics},
  \href{https://doi.org/10.1103/PhysRevB.99.224307}{\emph{Phys. Rev. B}
  {\bfseries 99} (Jun, 2019) 224307}.

\bibitem{jianMeasurementInducedPhaseTransition2021a}
S.-K. Jian, C.~Liu, X.~Chen, B.~Swingle and P.~Zhang,
  \emph{Measurement-{{Induced Phase Transition}} in the {{Monitored
  Sachdev-Ye-Kitaev Model}}},
  \href{https://doi.org/10.1103/PhysRevLett.127.140601}{\emph{Phys. Rev. Lett.}
  {\bfseries 127} (2021) 140601}.

\bibitem{jianMeasurementinducedCriticalityRandom2020a}
C.-M. Jian, Y.-Z. You, R.~Vasseur and A.~W.~W. Ludwig,
  \emph{Measurement-induced criticality in random quantum circuits},
  \href{https://doi.org/10.1103/PhysRevB.101.104302}{\emph{Phys. Rev. B}
  {\bfseries 101} (2020) 104302}.

\bibitem{zhangUniversalEntanglementTransitions2022}
P.~Zhang, C.~Liu, S.-K. Jian and X.~Chen, \emph{Universal {{Entanglement
  Transitions}} of {{Free Fermions}} with {{Long-range Non-unitary Dynamics}}},
  \href{https://doi.org/10.22331/q-2022-05-27-723}{\emph{Quantum} {\bfseries 6}
  (2022) 723}.

\bibitem{zhouGeneralizedLindbladMaster2022}
Y.-N. Zhou, \emph{{Generalized Lindblad Master Equation for Measurement-induced
  Phase Transition}},  \href{https://arxiv.org/abs/2204.09049}{{\ttfamily
  2204.09049}}.

\bibitem{Gross:MBL2016}
J.~yoon Choi, S.~Hild, J.~Zeiher, P.~Schauß, A.~Rubio-Abadal, T.~Yefsah
  et~al., \emph{Exploring the many-body localization transition in two
  dimensions}, \href{https://doi.org/10.1126/science.aaf8834}{\emph{Science}
  {\bfseries 352} (2016) 1547--1552}.

\bibitem{smithManybodyLocalizationQuantum2016}
J.~Smith, A.~Lee, P.~Richerme, B.~Neyenhuis, P.~W. Hess, P.~Hauke et~al.,
  \emph{Many-body localization in a quantum simulator with programmable random
  disorder}, \href{https://doi.org/10.1038/nphys3783}{\emph{Nat. Phys.}
  {\bfseries 12} (2016) 907--911}.

\bibitem{bernienProbingManybodyDynamics2017}
H.~Bernien, S.~Schwartz, A.~Keesling, H.~Levine, A.~Omran, H.~Pichler et~al.,
  \emph{Probing many-body dynamics on a 51-atom quantum simulator},
  \href{https://doi.org/10.1038/nature24622}{\emph{Nature} {\bfseries 551}
  (2017) 579--584}.

\bibitem{LaiManybodyHilbertSpace2023}
P.~Zhang, H.~Dong, Y.~Gao, L.~Zhao, J.~Hao, J.-Y. Desaules et~al.,
  \emph{Many-body {{Hilbert}} space scarring on a superconducting processor},
  \href{https://doi.org/10.1038/s41567-022-01784-9}{\emph{Nat. Phys.}
  {\bfseries 19} (Jan., 2023) 120--125}.

\bibitem{lindnerFloquetTopologicalInsulator2011}
N.~H. Lindner, G.~Refael and V.~Galitski, \emph{Floquet topological insulator
  in semiconductor quantum wells},
  \href{https://doi.org/10.1038/nphys1926}{\emph{Nature Phys} {\bfseries 7}
  (2011) 490--495}.

\bibitem{rechtsmanPhotonicFloquetTopological2013}
M.~C. Rechtsman, J.~M. Zeuner, Y.~Plotnik, Y.~Lumer, D.~Podolsky, F.~Dreisow
  et~al., \emph{Photonic {{Floquet}} topological insulators},
  \href{https://doi.org/10.1038/nature12066}{\emph{Nature} {\bfseries 496}
  (2013) 196--200}.

\bibitem{RoushanQuantumInformationPhases2023}
Google-Quantum-AI and Collaborators, \emph{Quantum information phases in
  space-time: Measurement-induced entanglement and teleportation on a noisy
  quantum processor},  \href{https://arxiv.org/abs/2303.04792}{{\ttfamily
  2303.04792}}.

\bibitem{Hazzard:2014bx}
K.~R.~A. Hazzard, B.~Gadway, M.~{Foss-Feig}, B.~Yan, S.~A. Moses, J.~P. Covey
  et~al., \emph{Many-{{Body Dynamics}} of {{Dipolar Molecules}} in an {{Optical
  Lattice}}}, \href{https://doi.org/10.1103/PhysRevLett.113.195302}{\emph{Phys.
  Rev. Lett.} {\bfseries 113} (2014) 195302}.

\bibitem{Yan:2013fn}
B.~Yan, S.~A. Moses, B.~Gadway, J.~P. Covey, K.~R.~A. Hazzard, A.~M. Rey
  et~al., \emph{Realizing a lattice spin model with polar molecules},
  \href{https://doi.org/10.1038/nature12483}{\emph{Nature} {\bfseries 501}
  (2013) 521--525}.

\bibitem{Lukin17NV}
J.~Choi, S.~Choi, G.~Kucsko, P.~C. Maurer, B.~J. Shields, H.~Sumiya et~al.,
  \emph{Depolarization dynamics in a strongly interacting solid-state spin
  ensemble}, \href{https://doi.org/10.1103/PhysRevLett.118.093601}{\emph{Phys.
  Rev. Lett.} {\bfseries 118} (Mar, 2017) 093601}.

\bibitem{Lukin18NV}
G.~Kucsko, S.~Choi, J.~Choi, P.~C. Maurer, H.~Zhou, R.~Landig et~al.,
  \emph{Critical thermalization of a disordered dipolar spin system in
  diamond}, \href{https://doi.org/10.1103/PhysRevLett.121.023601}{\emph{Phys.
  Rev. Lett.} {\bfseries 121} (Jul, 2018) 023601}.

\bibitem{smaleObservationTransitionDynamical2019}
S.~Smale, P.~He, B.~A. Olsen, K.~G. Jackson, H.~Sharum, S.~Trotzky et~al.,
  \emph{Observation of a transition between dynamical phases in a quantum
  degenerate {{Fermi}} gas},
  \href{https://doi.org/10.1126/sciadv.aax1568}{\emph{Sci. Adv.} {\bfseries 5}
  (2019) eaax1568}.

\bibitem{Bloch:2012ee}
P.~Schau{\ss}, M.~Cheneau, M.~Endres, T.~Fukuhara, S.~Hild, A.~Omran et~al.,
  \emph{Observation of spatially ordered structures in a two-dimensional
  {{Rydberg}} gas}, \href{https://doi.org/10.1038/nature11596}{\emph{Nature}
  {\bfseries 491} (2012) 87--91}.

\bibitem{Signoles:2019us}
A.~Signoles, T.~Franz, R.~Ferracini~Alves, M.~G{\"a}rttner, S.~Whitlock,
  G.~Z{\"u}rn et~al., \emph{Glassy {{Dynamics}} in a {{Disordered Heisenberg
  Quantum Spin System}}},
  \href{https://doi.org/10.1103/PhysRevX.11.011011}{\emph{Phys. Rev. X}
  {\bfseries 11} (2021) 011011}.

\bibitem{gabardosRelaxationCollectiveMagnetization2020}
L.~Gabardos, B.~Zhu, S.~Lepoutre, A.~M. Rey, B.~{Laburthe-Tolra} and L.~Vernac,
  \emph{Relaxation of the {{Collective Magnetization}} of a {{Dense 3D Array}}
  of {{Interacting Dipolar}} \${{S}}=3\$ {{Atoms}}},
  \href{https://doi.org/10.1103/PhysRevLett.125.143401}{\emph{Phys. Rev. Lett.}
  {\bfseries 125} (2020) 143401}.

\bibitem{Suter06}
H.~G. Krojanski and D.~Suter, \emph{Reduced decoherence in large quantum
  registers}, \href{https://doi.org/10.1103/PhysRevLett.97.150503}{\emph{Phys.
  Rev. Lett.} {\bfseries 97} (Oct, 2006) 150503}.

\bibitem{Suter07}
M.~Lovri\ifmmode~\acute{c}\else \'{c}\fi{}, H.~G. Krojanski and D.~Suter,
  \emph{Decoherence in large quantum registers under variable interaction with
  the environment},
  \href{https://doi.org/10.1103/PhysRevA.75.042305}{\emph{Phys. Rev. A}
  {\bfseries 75} (Apr, 2007) 042305}.

\bibitem{kamenev2023field}
A.~Kamenev, \emph{Field theory of non-equilibrium systems}.
\newblock Cambridge University Press, 2023.

\bibitem{stefanucci2013nonequilibrium}
G.~Stefanucci and R.~van Leeuwen, \emph{Nonequilibrium Many-Body Theory of
  Quantum Systems: A Modern Introduction}.
\newblock Cambridge University Press, 2013.

\bibitem{kitaevalexei2015}
A.~Kitaev, \emph{Talk given at kitp program: Entanglement in
  strongly-correlated quantum matter},  2015.

\bibitem{maldacenaRemarksSachdevYeKitaevModel2016a}
J.~Maldacena and D.~Stanford, \emph{Remarks on the {{Sachdev-Ye-Kitaev}}
  model}, \href{https://doi.org/10.1103/PhysRevD.94.106002}{\emph{Phys. Rev. D}
  {\bfseries 94} (2016) 106002}.

\bibitem{songStronglyCorrelatedMetal2017a}
X.-Y. Song, C.-M. Jian and L.~Balents, \emph{Strongly {{Correlated Metal
  Built}} from {{Sachdev-Ye-Kitaev Models}}},
  \href{https://doi.org/10.1103/PhysRevLett.119.216601}{\emph{Phys. Rev. Lett.}
  {\bfseries 119} (2017) 216601}.

\bibitem{davisonThermoelectricTransportDisordered2017}
R.~A. Davison, W.~Fu, A.~Georges, Y.~Gu, K.~Jensen and S.~Sachdev,
  \emph{Thermoelectric transport in disordered metals without quasiparticles:
  {{The Sachdev-Ye-Kitaev}} models and holography},
  \href{https://doi.org/10.1103/PhysRevB.95.155131}{\emph{Phys. Rev. B}
  {\bfseries 95} (2017) 155131}.

\bibitem{guoTransportChaosLattice2019b}
H.~Guo, Y.~Gu and S.~Sachdev, \emph{Transport and chaos in lattice
  sachdev-ye-kitaev models},
  \href{https://doi.org/10.1103/PhysRevB.100.045140}{\emph{Phys. Rev. B}
  {\bfseries 100} (Jul, 2019) 045140}.

\bibitem{Gu:2020it}
Y.~Gu, A.~Kitaev, S.~Sachdev and G.~Tarnopolsky, \emph{Notes on the complex
  {{Sachdev-Ye-Kitaev}} model},
  \href{https://doi.org/10.1007/JHEP02(2020)157}{\emph{J. High Energ. Phys.}
  {\bfseries 02} (2020) 157}.

\bibitem{zhouDisconnectingTraversableWormhole2021}
T.-G. Zhou, L.~Pan, Y.~Chen, P.~Zhang and H.~Zhai, \emph{Disconnecting a
  traversable wormhole: {{Universal}} quench dynamics in random spin models},
  \href{https://doi.org/10.1103/PhysRevResearch.3.L022024}{\emph{Phys. Rev.
  Research} {\bfseries 3} (2021) L022024}.

\bibitem{sachdevGaplessSpinfluidGround1993}
S.~Sachdev and J.~Ye, \emph{Gapless spin-fluid ground state in a random quantum
  {{Heisenberg}} magnet},
  \href{https://doi.org/10.1103/PhysRevLett.70.3339}{\emph{Phys. Rev. Lett.}
  {\bfseries 70} (1993) 3339--3342}.

\bibitem{Baldwin:2019dki}
C.~L. Baldwin and B.~Swingle, \emph{{Quenched vs Annealed: Glassiness from SK
  to SYK}}, \href{https://doi.org/10.1103/PhysRevX.10.031026}{\emph{Phys. Rev.
  X} {\bfseries 10} (2020) 031026}.

\bibitem{zhangObstacleSubAdSHolography2021}
P.~Zhang, Y.~Gu and A.~Kitaev, \emph{An obstacle to sub-{{AdS}} holography for
  {{SYK-like}} models}, \href{https://doi.org/10.1007/JHEP03(2021)094}{\emph{J.
  High Energ. Phys.} {\bfseries 03} (2021) 094}.

\bibitem{Zhang:2023wtr}
R.~Zhang and H.~Zhai, \emph{{Universal hypothesis of autocorrelation function
  from Krylov complexity}},
  \href{https://doi.org/10.1007/s44214-024-00054-4}{\emph{Quant. Front.}
  {\bfseries 3} (2024) 7}, [\href{https://arxiv.org/abs/2305.02356}{{\ttfamily
  2305.02356}}].

\end{thebibliography}\endgroup

\end{document}